\documentclass[conference,a4paper,10pt]{IEEEtran}

\usepackage[
    unicode=true,
    pdftoolbar=true,
    pdfmenubar=true,
    pdffitwindow=true,
    pdfstartview={FitH},
    pdftitle={Estimation of Radio Channel Parameters in Case of an Unknown Transmitter},
    pdfauthor={Häfner},
    pdfnewwindow=true,
    breaklinks=true,
    hidelinks
]{hyperref}

\usepackage{graphicx}
\usepackage{pstool}
\usepackage{ctable}
\usepackage{lettrine}
\usepackage{xcomment}
\usepackage{times}
\usepackage{amsmath,amsfonts,amssymb,amsxtra}
\usepackage [utf8x]{inputenc}
\usepackage[T1]{fontenc}
\usepackage{multirow}
\usepackage{cite}
\usepackage{color}
\usepackage{placeins}
\usepackage{rotating}
\usepackage{trfsigns}

\newcommand{\slfrac}[2]{\left. #1 \middle/ #2 \right.}

\usepackage{tikz}
\usetikzlibrary{automata,arrows,shadows,shapes,decorations.pathmorphing,backgrounds,positioning,fit,shapes.multipart,patterns,mindmap,trees,chains,calc,matrix}
\definecolor{tui-green}{rgb}{0,0.455,0.478}
\definecolor{tui-blue}{rgb}{0,0.2,0.349}
\definecolor{tui-orange}{rgb}{1.0,0.475,0}
\definecolor{tui-lightblue}{rgb}{0.706,0.863,0.863}

\usepackage[load-configurations=abbreviations,load-configurations=binary]{siunitx}	
\DeclareSIUnit{\mms}{\milli\squaremetre}
\DeclareSIUnit{\inch}{in}
\DeclareSIUnit{\inchs}{in\squared}
\DeclareSIUnit{\mil}{mil}
\DeclareSIUnit{\Msps}{Msps}
\DeclareSIUnit{\Mbps}{Mbps}
\DeclareSIUnit{\LSB}{LSB}
\DeclareSIUnit{\pFS}{\percent FS}
\DeclareSIUnit{\dBc}{\deci\bel c}
\DeclareSIUnit{\dBm}{\deci\bel m}
\DeclareSIUnit{\dBFS}{\deci\bel FS}
\DeclareSIUnit{\dB}{\deci\bel}
\DeclareSIUnit{\dBi}{\deci\bel i}
\DeclareSIUnit{\hex}{0x}
\DeclareSIUnit{\vp}{\volt_{\text{p}}}
\DeclareSIUnit{\vpp}{\volt_{\text{pp}}}
\DeclareSIUnit{\kb}{\kilo\bit}
\DeclareSIUnit{\kB}{\kilo\byte}
\DeclareSIUnit{\MB}{\mega\byte}
\DeclareSIUnit{\GHz}{\giga\hertz}
\DeclareSIUnit{\MHz}{\mega\hertz}

\DeclareSIUnit{\mus}{\micro\second}
\DeclareSIUnit{\ns}{\nano\second}
\DeclareSIUnit{\fs}{\femto\second}

\title{Estimation of Radio Channel Parameters in Case of an Unknown Transmitter}%

\author{
\IEEEauthorblockN{Stephan Häfner and Reiner Thomä}\\
\IEEEauthorblockA{Electronic Measurement Research Lab\\ Ilmenau University of Technology, Germany\\stephan.haefner@tu-ilmenau.de}
}

\begin{document}


\maketitle

\begin{abstract}
This paper investigates the estimation of radio channel parameters from receiver data, whereby the transmitter is fully unknown. We use a multipath model to describe the radio channel between transmitter and receiver. According to this model, we discuss the accessibility of parameters for estimation. Based on the Maximum-Likelihood principle, we derive a cost function. A second cost function is derived from the cross relation between the receiver channels. To estimate the parameters, we seek for the minimum of these cost functions. The performance of the presented cost functions are compared in simulations.
\end{abstract}
\vspace{3mm}
\textbf{Keywords --} Radio channel parameters, Parameter estimation, Maximum-Likelihood principle, Channel-Cross-Relation

\section{Introduction}
\lettrine[lines=2, lhang=0, loversize=0]{E}{stimation} of model parameters is a task of wide interest in engineering applications. In channel sounding, measurement data are used to estimate parameters of a radio channel impulse response model. Here, the signal at transmitter and receiver are known, such that the impulse response can be estimated by deconvolution. Algorithms for parameter estimation based on radio channel impulse responses are known (JADE \cite{Vanderveen97}, RIMAX \cite{Richter05}).

In some scenarios, the transmitter signal is unknown. This can happen, if the transmitter acts as a jammer. So, methods of blind channel estimation are necessary. Such methods are described in the literature (e.g. \cite{Xu95}, \cite{Hua96}, \cite{Moulines95}). After the blind channel estimation step, an algorithm for parameter estimation can be applied. This two step approach is harmful, because two estimation methods are needed. In this paper we describe an approach, which estimates the channel parameters directly from the measured receiver data. We derive two cost functions, which depends only on the channel parameters. Estimation is done by minimisation of the cost functions with the Levenberg-Marquardt algorithm. These estimated parameters can be used to locate the transmitter (see \cite{Algeier10}).

The paper is organized as follows: the signal model and a discussion on the accessibility of the model parameters is given in the next Section. In Section \ref{sec:CML} a cost function based on the Maximum-Likelihood principle for an unknown transmitter signal is derived. A second cost function, which uses the relation between the receiver channels, is presented in section \ref{sec:CCR}. The simulation results and final conclusions can be found in the Sections \ref{sec:Simulations} and \ref{sec:Conclusion}.

We use standard notation in this paper. Matrices (in capital letters) and vectors are in boldface. We define the matrix operations $(.)^T$, $(.)^H$, $(.)^+$ as the Transpose, Hermitian and Pseudo Inverse of a matrix, respectively. Symbol $\diamond$ represents the Khatri-Rao product. The second norm of a vector is stated as $\left| \left| . \right| \right|_2$.

\section{Radio Channel Model}
\label{sec:Radio Channel Model}

The noiseless receiver signal in the frequency domain for a specular propagation path can be modelled by a ray-optical model \cite[pp. 10]{Richter05}:

\begin{align}
	\nonumber
	x(f) = \begin{bmatrix} b_{H}^{Rx}(\phi^{Rx},\theta^{Rx}) & b_{V}^{Rx}(\phi^{Rx},\theta^{Rx}) \end{bmatrix} \cdot \begin{bmatrix} \gamma_{HH} & \gamma_{HV} \\ \gamma_{VH} & \gamma_{VV} \end{bmatrix} \\
	\cdot \begin{bmatrix} b_{H}^{Tx}(\phi^{Tx},\theta^{Tx}) \\ b_{V}^{Tx}(\phi^{Tx},\theta^{Tx}) \end{bmatrix} \cdot e^{-j 2 \pi f \tau} \cdot s'(f) \cdot G^{Rx}(f) \cdot G^{Tx}(f)
\end{align}

where

\begin{itemize}
	\item $b_{H/V}^{Rx}(\phi^{Rx},\theta^{Rx})$ angles-of-arrival dependent complex beam pattern of the receiver antenna for horizontal/vertical polarisation
	\item $b_{H/V}^{Tx}(\phi^{Tx},\theta^{Tx})$ angles-of-departure dependent complex beam pattern of the transmit antenna for horizontal/vertical polarisation
	\item $G^{Tx}(f)$ transmitter frequency response
	\item $G^{Rx}(f)$ receiver frequency response
	\item $s'(f)$ transmitter signal 
	\item $e^{-j 2 \pi f \tau}$ complex exponential for delay $\tau$
	\item $\footnotesize \begin{bmatrix} \gamma_{HH} & \gamma_{HV} \\ \gamma_{VH} & \gamma_{VV} \end{bmatrix}$ matrix of polarimetric transmission coefficients
\end{itemize}

and $\phi^{Rx},\theta^{Rx}$ are the azimuth (AoA) and elevation (EoA) angle-of-arrival and $\phi^{Tx},\theta^{Tx}$ are the azimuth and elevation angle-of-departure, respectively.

We assume no information about the transmitter. Therefore, we cannot distinguish between the transmitter signal and the transmitter frequency response. Furthermore, we cannot estimate the matrix of polarimetric transmission coefficients, because the transmit antenna beam pattern is unknown. To overcome this issues, we combine parameters:

\begin{align}
	& \begin{bmatrix} \gamma_{H} \\ \gamma_{V} \end{bmatrix} = \begin{bmatrix} \gamma_{HH} & \gamma_{HV} \\ \gamma_{VH} & \gamma_{VV} \end{bmatrix} \cdot \begin{bmatrix} b_{H}^{Tx}(\phi^{Tx},\theta^{Tx}) \\ b_{V}^{Tx}(\phi^{Tx},\theta^{Tx}) \end{bmatrix} \\
	& s(f) = s'(f) \cdot G^{Tx}(f)
\end{align}

Here, $\begin{bmatrix} \gamma_{H} & \gamma_{V} \end{bmatrix}^T$ describes the polarimetric path weights at the receiver and $s(f)$ is now denoted as the transmitter signal. Furthermore, we assume a calibrated receiver system, such that $G^{Rx}(f)=1$. With this simplifications we get:

\begin{align}
	x(f) = \begin{bmatrix} b_{H}^{Rx}(\phi^{Rx},\theta^{Rx}) \\ b_{V}^{Rx}(\phi^{Rx},\theta^{Rx}) \end{bmatrix}^T \cdot \begin{bmatrix} \gamma_{H} \\ \gamma_{V} \end{bmatrix} \cdot e^{-j 2 \pi f \tau} \cdot s(f)
\end{align}

For notational convenience, we refer $\phi^{Rx}$ and $\theta^{Rx}$ now to $\phi$ and $\theta$, respectively. 
We extend this model to a SIMO-model, because an antenna array at the receiver and a single transmitter with one antenna are assumed. Furthermore, we extend the model to the multipath case. For $P$ propagation paths and $M_R$ receiver antennas, we get the model described in \cite{Yang94}:

\begin{align}
	\mathbf{x}(f) = \mathbf{B}(\boldsymbol{\phi},\boldsymbol{\theta}) \cdot \boldsymbol{\Gamma}(\boldsymbol{\gamma_H},\boldsymbol{\gamma_V}) \cdot \mathbf{e}(\boldsymbol{\tau},f) \cdot s(f)
\end{align}

where

\begin{itemize}
	\item $\mathbf{B}(\boldsymbol{\phi},\boldsymbol{\theta}) \in \mathbb{C}^{M_R \times 2P}$ polarimetric steering matrix
	\item $\boldsymbol{\Gamma}(\boldsymbol{\gamma_H},\boldsymbol{\gamma_V}) \in \mathbb{C}^{2P \times 2P}$ diagonal matrix with polarimetric path weights
	\item $\mathbf{e}(\boldsymbol{\tau},f) \in \mathbb{C}^{2P \times 1}$ vector of complex exponentials for vertical and horizontal polarisation
\end{itemize}

$K$ samples are measured at each receiver antenna port. The extended model is then given by:

\begin{align}
	\mathbf{X} = \mathbf{B}(\boldsymbol{\phi},\boldsymbol{\theta}) \cdot \boldsymbol{\Gamma}(\boldsymbol{\gamma_H},\boldsymbol{\gamma_V}) \cdot \mathbf{E}(\boldsymbol{\tau}) \cdot \mathbf{S}(\mathbf{s}) = \mathbf{H}(\boldsymbol{\alpha}) \cdot \mathbf{S}(\mathbf{s})
	\label{eq: model}
\end{align}

with $\mathbf{E}(\boldsymbol{\tau}) \in \mathbb{C}^{2P \times K}$ containing the vectors of complex exponentials for each frequency bin and $\mathbf{S}(\mathbf{s}) \in \mathbb{C}^{K \times K}$ the diagonal matrix of the transmitter signal vector. For simplification we introduced the vector of path parameters $\boldsymbol{\alpha} = \begin{bmatrix} \boldsymbol{\phi}^T & \boldsymbol{\theta}^T & \boldsymbol{\gamma_H}^T & \boldsymbol{\gamma_V}^T & \boldsymbol{\tau}^T \end{bmatrix}^T$ and the channel matrix $\mathbf{H}(\boldsymbol{\alpha})$.

According to equation \eqref{eq: model}, we discuss the accessibility of the model parameters. First, there is no synchronisation between transmitter and receiver. Hence absolute delays are not accessible and only relative delays can be estimated, what is also stated in \cite{Gunther96}. Second, only relative path weights can be estimated. Therefore, we refer each propagation path to the earliest arrival.

To complete the signal model, additive Gaussian noise (modelling the measurement noise and the model error) and no dense multipath components are assumed at the receiver:

\begin{align}
	\mathbf{Y} = \mathbf{X}(\boldsymbol{\alpha},\mathbf{s}) + \mathbf{N}
	\label{eq: Rxmodel}
\end{align}

Algorithms to estimate radio channel parameters are widely known. For arbitrary antenna geometries, only MUSIC or Maximum-Likelihood methods are usable. The advantage of Maximum-Likelihood methods is, that all parameters can be estimated jointly. Furthermore, only one optimum have to be detected in Maximum-Likelihood methods. In the MUSIC method we have to search for $P$ peaks, which is more difficult. Hence, only Maximum-Likelihood methods are taken into account.

We are only interested in the path parameters, whereas the transmitter signal is considered as a nuisance parameter. Therefore, a cost function independent of the transmitter signal is needed for an optimisation based parameter estimation procedure. In the following sections, two cost functions are derived to overcome this requirement.

Throughout the rest of the paper the number $P$ of propagation paths is assumed as known.
\section{Constrained-Maximum-Likelihood cost function}
\label{sec:CML}

In the first cost function the unknown transmitter signal is replaced by an estimator. For that purpose we assume the transmitter signal as deterministic.

To derive an estimator, we explore the structure of signal matrix $\mathbf{S}$ in equation \eqref{eq: model}. We remember, that the signal matrix has a diagonal structure. Hence, only the diagonal elements have to be estimated. To exploit this fact, we restate the model in \eqref{eq: model} using the $\text{vec}\{.\}$ operator, which stacks the columns of a matrix:

\begin{align}
	\nonumber
	& \text{vec} \{ \mathbf{Y} \} = \mathbf{y} = \left( \mathbf{I}_K \diamond \mathbf{H}(\boldsymbol{\alpha}) \right) \cdot \mathbf{s} + \text{vec} \{ \mathbf{N} \} \\
	& = \mathbf{\tilde{H}}(\boldsymbol{\alpha}) \cdot \mathbf{s} + \mathbf{n}
	\label{eq: CMLmodel}
\end{align}

with $\mathbf{I}_K$ the unity matrix of size $K$. According to this equation, we can develop an estimator for the signal vector $\mathbf{s}$.

Based on the Maximum-Likelihood principle and the Gaussian noise assumption, the following probability-density-function describes an observation based on equation \eqref{eq: CMLmodel}: 

\begin{align}
	p(\mathbf{y},\boldsymbol{\alpha}) = \frac{ e^{- \left(\mathbf{y}-\mathbf{\tilde{H}}(\boldsymbol{\alpha}) \cdot \mathbf{s}\right)^H \cdot \mathbf{\tilde{R}}^{-1} \cdot \left(\mathbf{y}-\mathbf{\tilde{H}}(\boldsymbol{\alpha}) \cdot \mathbf{s}\right) } }{\pi^{M_RK} \cdot det(\mathbf{\tilde{R}})}
\end{align}

with $\mathbf{\tilde{R}}$ the noise covariance matrix. Typically, the negative log-Likelihood function is used as cost function:

\begin{align}
	\nonumber
	& -\text{ln} \left(p(\mathbf{y},\boldsymbol{\alpha})\right) = M_RK \text{ln}(\pi) + \text{ln} \left(det(\mathbf{\tilde{R}})\right) \\
	& + \left(\mathbf{y}-\mathbf{\tilde{H}}(\boldsymbol{\alpha}) \cdot \mathbf{s}\right)^H \cdot \mathbf{\tilde{R}}^{-1} \cdot \left(\mathbf{y}-\mathbf{\tilde{H}}(\boldsymbol{\alpha}) \cdot \mathbf{s}\right)
	\label{eq: MLCost}
\end{align}

An estimator for the signal vector in equation \eqref{eq: MLCost} is the Best-Linear-Unbiased-Estimator (BLUE):

\begin{align}
	\mathbf{\hat{s}} = \left( \mathbf{\tilde{H}}(\boldsymbol{\alpha})^H \cdot \mathbf{\tilde{R}}^{-1} \cdot \mathbf{\tilde{H}}(\boldsymbol{\alpha}) \right)^{-1} \cdot \mathbf{\tilde{H}}(\boldsymbol{\alpha})^H \cdot \mathbf{\tilde{R}}^{-1} \cdot \mathbf{y}
	\label{eq: BLUE}
\end{align}

For simplification, we use the Cholesky decomposition $\mathbf{\tilde{R}}^{-1} = \mathbf{\tilde{L}}^{-1^H} \cdot \mathbf{\tilde{L}}^{-1}$ of the noise covariance matrix and introduce the abbreviations: 

\begin{align}
	& \mathbf{\tilde{H}}_L = \mathbf{\tilde{L}}^{-1} \cdot \mathbf{\tilde{H}} \\
	& \mathbf{y}_L = \mathbf{\tilde{L}}^{-1} \cdot \mathbf{y}
\end{align}

Using this abbreviations and inserting \eqref{eq: BLUE} in \eqref{eq: MLCost}, we get the Constrained-Maximum-Likelihood (CML) cost function w.r.t. the path parameters:

\begin{align}
	C_{CML}(\boldsymbol{\alpha}) = \left| \left| \mathbf{y}_L - \mathbf{\tilde{H}}_L(\boldsymbol{\alpha}) \cdot \mathbf{\tilde{H}}_L(\boldsymbol{\alpha})^+ \cdot \mathbf{y}_L \right| \right|_2^2
	\label{eq: CML}
\end{align}
\section{Channel-Cross-Relation cost function}
\label{sec:CCR}

The second cost function is an extension of the idea described in \cite{Xu95} to a frequency domain parametric channel model like \eqref{eq: model}. In case of a SIMO channel, the following relation between two arbitrary receiver channels in the frequency domain exists:

\begin{align}
	\underbrace{h^{(i)}(k) \cdot s(k)}_{= x^{(i)}(k)} \cdot h^{(j)}(k) - \underbrace{h^{(j)}(k) \cdot s(k)}_{= x^{(j)}(k)} \cdot h^{(i)}(k) = 0
\end{align}

This relation is true, if no noise occurs in the receiver and the radio channel behaves exactly like the model assumption. Furthermore, the relation is independent of the transmitter signal. 

We extend this relation to all receiver channel combinations using the Data Selection Transform $DST \{.\}$ described in \cite{Zeng95}.

\begin{align}
	\nonumber
	& DST \left\{ \mathbf{x}(k) \right\} \cdot \mathbf{h}(k) \\
	& = DST \left\{ \mathbf{x}(k) \right\} \left(\mathbf{e}(k,\boldsymbol{\tau})^T \diamond \mathbf{B}(\boldsymbol{\phi}, \boldsymbol{\theta}) \right) \cdot \boldsymbol{\gamma} = \mathbf{0}
\end{align}

The above equation describes the relation between every receiver channel at one frequency bin. Hence, vector $\mathbf{h}(k)$ is a column of the channel matrix in \eqref{eq: model}. To extend this relation to all measured frequency bins, we introduce the Diagonal Data Selection Transform $DDST \{.\}$:

\begin{align}
	& DDST \left\{ \mathbf{X} \right\} = 
	\begin{bmatrix}
	DST \left\{ \mathbf{x}(1) \right\} & \dots &  \mathbf{0} \\ 
	\vdots & \ddots &  \vdots \\ 
	\mathbf{0} & \dots & DST \left\{ \mathbf{x}(K) \right\}
	\end{bmatrix} 
\end{align}

According to this transformation and the relation $\mathbf{h} = \begin{bmatrix} \mathbf{h}(1)^T & \dots & \mathbf{h}(K)^T \end{bmatrix}^T = \text{vec}\{\mathbf{H}\}$, we can write

\begin{align}
	\nonumber
	& DDST \left\{ \mathbf{X} \right\} \cdot \mathbf{h} \\
	& = DDST \left\{ \mathbf{X} \right\} \cdot \left( \mathbf{E}(\boldsymbol{\tau})^T \diamond \mathbf{B}(\boldsymbol{\phi}, \boldsymbol{\theta}) \right) \cdot \boldsymbol{\gamma} = \mathbf{0}
	\label{eq: CCR_Relation}
\end{align}

In case of measurement noise and model errors, the equality with the zero vector in \eqref{eq: CCR_Relation} is not given. According to the Maximum-Likelihood principle and under the Gaussian noise assumption, we get the squared second norm of equation \eqref{eq: CCR_Relation} as cost function:

\begin{align}
	\left| \left| DDST \left\{ \mathbf{Y} \right\} \cdot \left( \mathbf{E}(\boldsymbol{\tau})^T \diamond \mathbf{B}(\boldsymbol{\phi},\boldsymbol{\theta}) \right) \cdot \boldsymbol{\gamma} \right| \right|_2^2
	\label{eq: uCCR}
\end{align}

Swindlehurst and Kailath showed in \cite{Swindlehurst92} that the performance of the MUSIC cost function could be improved by weighting, if non-uniform errors occur. Hence, we introduce a weighting of the cost function \eqref{eq: uCCR} and get the Channel-Cross-Relation (CCR) cost function w.r.t. the path parameters:

\begin{align}
	C_{CCR}(\boldsymbol{\alpha}) = \frac{\left| \left| DDST \left\{ \mathbf{Y} \right\} \cdot \left( \mathbf{E}(\boldsymbol{\tau})^T \diamond \mathbf{B}(\boldsymbol{\phi},\boldsymbol{\theta}) \right) \cdot \boldsymbol{\gamma} \right| \right|_2^2}{\left| \left| \left( \mathbf{E}(\boldsymbol{\tau})^T \diamond \mathbf{B}(\boldsymbol{\phi},\boldsymbol{\theta}) \right) \cdot \boldsymbol{\gamma} \right| \right|_2^2}
	\label{eq: CCR}
\end{align}

The CCR cost function has two advantages compared to the CML cost function. First, no Pseudo Inverse is needed. Only the Diagonal Data Selection Transform have to be computed once. That reduces computational complexity in an iterative optimisation procedure. Second, the path weights occur as linear parameters. Therefore, we can divide the optimisation into two steps, one step for the non-linear parameters (angle of arrival, delay) and one step for the linear parameters (path weights). Thus, the search space for the non-linear optimisation procedure is much smaller, what furthermore reduces computational complexity.

A disadvantage of the CCR cost function is her complexity, if derivatives are needed. For a gradient based optimisation procedure, partial derivatives of the cost function have to be calculated, which are much simpler for the CML cost function.
\section{Simulations}
\label{sec:Simulations}

\begin{figure}
	\centering
	\includegraphics[width=0.8\linewidth]{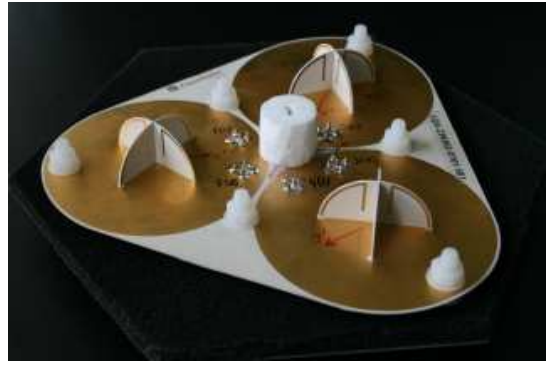}
	\caption{dual polarimetric L-Quad antenna array at Rx side}
	\label{fig:LQuad}
\end{figure}

To compare the proposed cost functions, Monte-Carlo simulations were conducted. For fixed channel parameters and a variable Signal-to-Noise-Ratio (SNR), we generated data samples according to our model \eqref{eq: Rxmodel}. As antenna at the receiver, we used the array shown in figure \ref{fig:LQuad}. The array consists of 3 spatial distributed sensors, whereas each sensor has a port for right-hand- and left-hand-circular polarisation. 
We assumed a band limited rectangular impulse as transmitter signal.

From the generated data, we estimated the parameters via minimisation of the cost functions. We used the Levenberg-Marquardt algorithm \cite{More77} as gradient-based optimiser. To calculate the derivatives of the steering matrix w.r.t. the angles, we utilised the EADF approach described in \cite{Landmann07}. This approach uses polarimetric calibration data to describe the complex antenna pattern.

For a estimated parameter set, we calculated the squared error for the azimuth- and elevation-angle of arrival, the normalised delay and the normalised path power defined as:

\begin{align}
	\slfrac{\left| \left| \begin{bmatrix} \gamma_{H}^{(p)} \\ \gamma_{V}^{(p)} \end{bmatrix} \right| \right|_2^2}{\left| \left| \begin{bmatrix} \gamma_{H}^{(1)} \\ \gamma_{V}^{(1)} \end{bmatrix} \right| \right|_2^2}
\end{align}

Squared errors were averaged over some trials and the square root was taken, to get the Root-Mean-Square-Error (RMSE) as performance measure.

We considered a 3 path scenario with the parameter values according to table \ref{tab:PathParameters}. The Signal-to-Noise-Ratio was varied from 0 dB up to 20 dB in 2 dB steps. Per SNR step, 1000 trials were generated and parameter estimation was done. 

\begin{figure}
	\centering
	\psfragfig[width=1\linewidth]{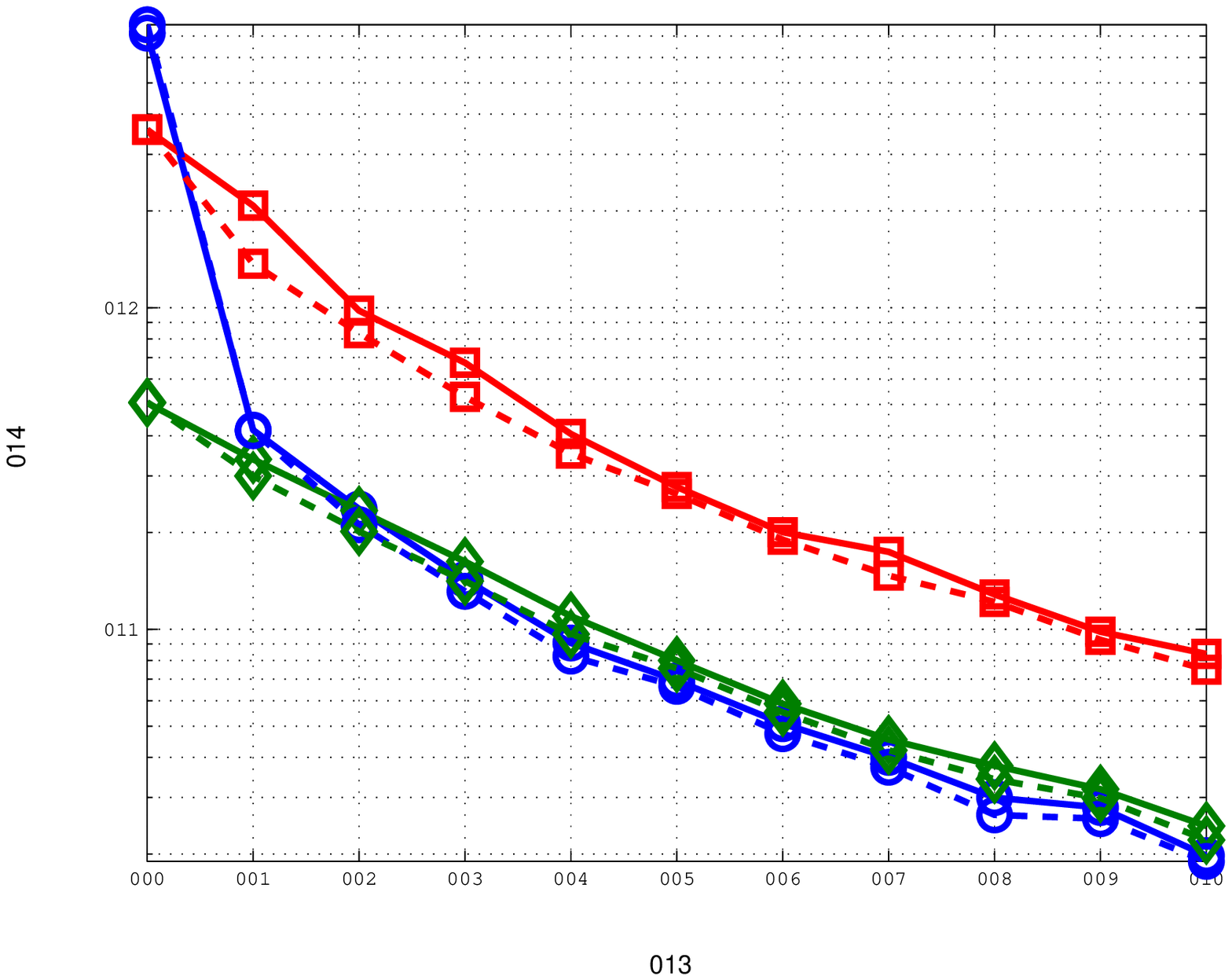}
	\caption{Root-Mean-Squared-Error of the azimuth of arrival for CCR cost function (solid line) and the CML cost function (dashed line), and the path number according to the marker type: first path ($\circ$), second path ($\Diamond$), third path ($\square$).}
	\label{fig:RMSE_AoA}
\end{figure}

\begin{figure}
	\centering
	\psfragfig[width=1\linewidth]{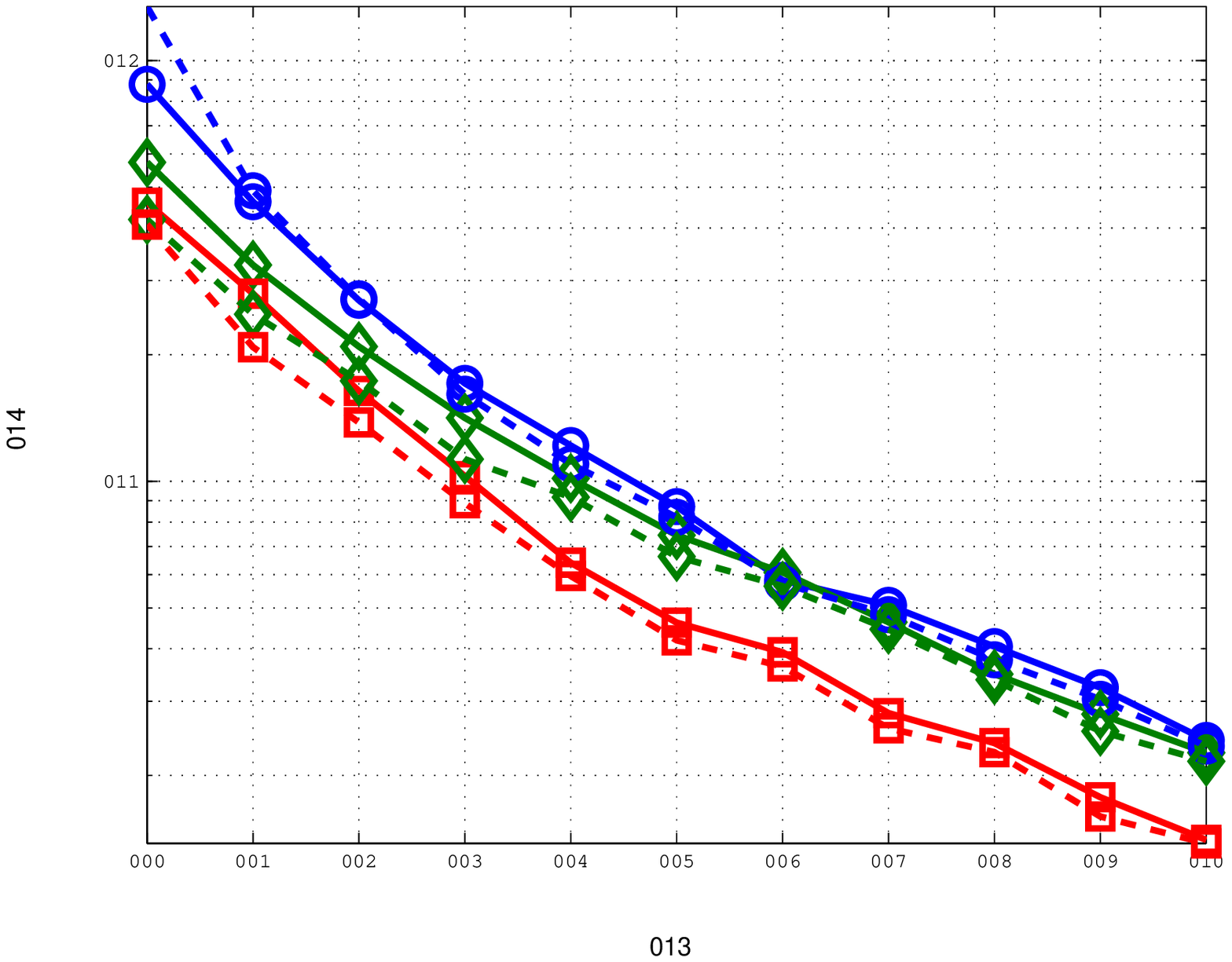}
	\caption{Root-Mean-Squared-Error of the elevation of arrival for CCR cost function (solid line) and the CML cost function (dashed line), and the path number according to the marker type: first path ($\circ$) and second path ($\Diamond$), third path ($\square$).}
	\label{fig:RMSE_EoA}
\end{figure}

\begin{figure}
	\centering
	\psfragfig[width=1\linewidth]{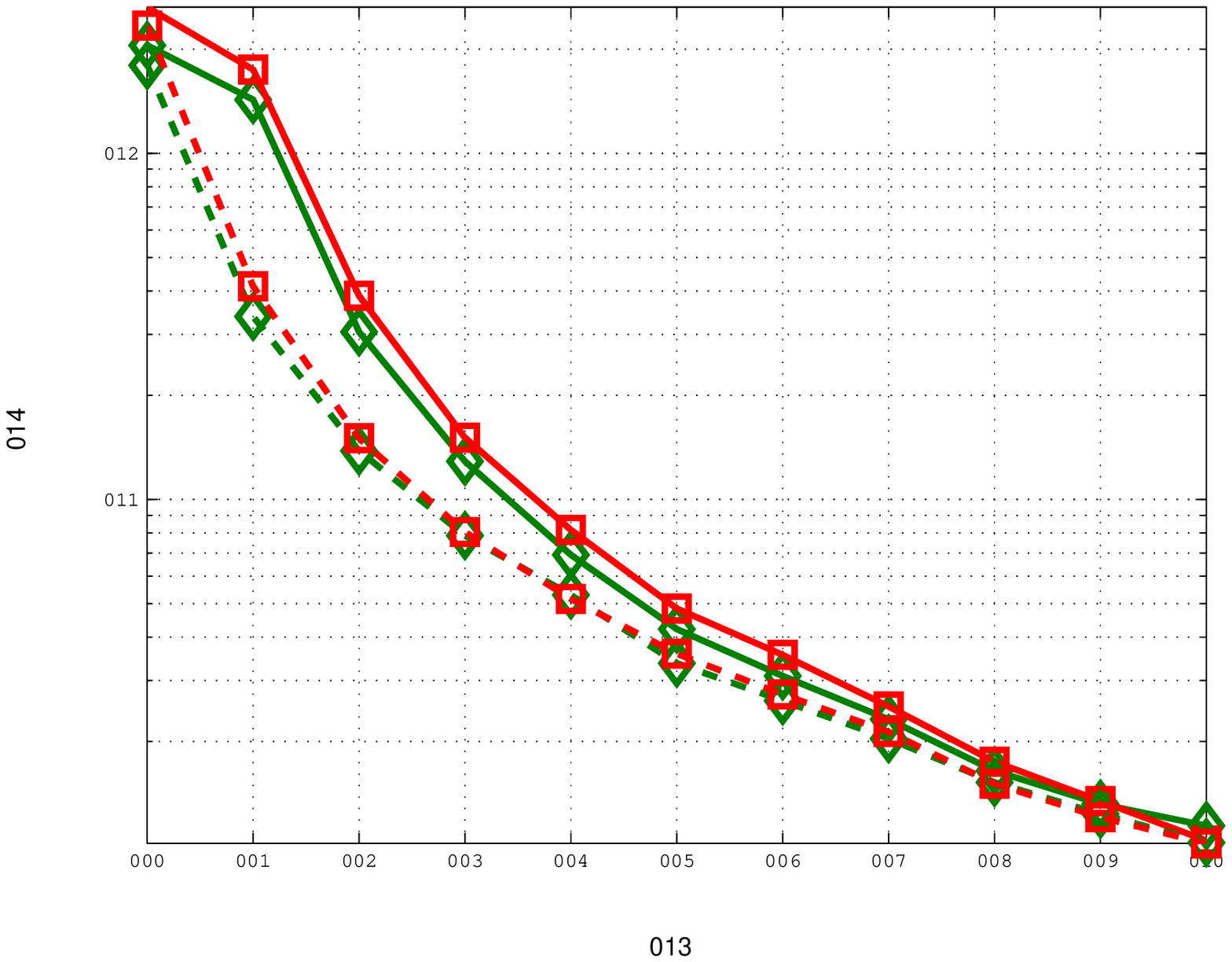}
	\caption{Root-Mean-Squared-Error of the normalised delay for CCR cost function (solid line) and the CML cost function (dashed line), and the path number according to the marker type: second path ($\Diamond$), third path ($\square$). Path number one is the reference path.}
	\label{fig:RMSE_Delay}
\end{figure}

\begin{figure}
	\centering
	\psfragfig[width=1\linewidth]{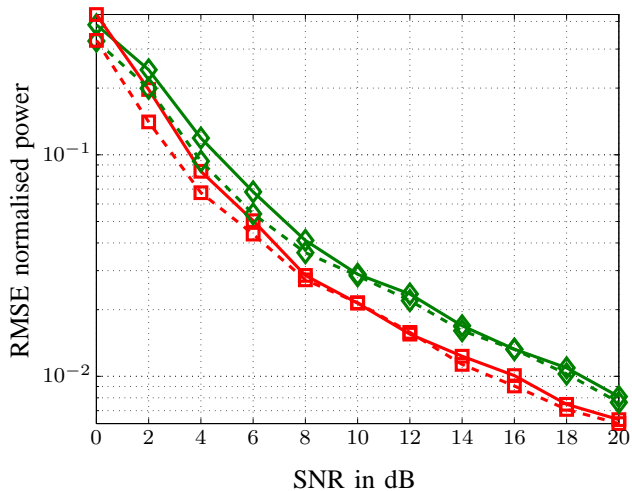}
	\caption{Root-Mean-Squared-Error of the normalised path power for CCR cost function (solid line) and the CML cost function (dashed line), and the path number according to the marker type: second path ($\Diamond$), third path ($\square$). Path number one is the reference path.}
	\label{fig:RMSE_Power}
\end{figure}

The curvatures of the RMSEs over the SNR are plotted in figures \ref{fig:RMSE_AoA}-\ref{fig:RMSE_Power}. The solid line represents the RMSE for the CCR cost function, whereas the dashed line represents the RMSE for the CML cost function. Marker types represent the path number: first path ($\circ$), second path ($\Diamond$), third path ($\square$).

\begin{table}[tbp]
	\centering
	\begin{tabular}{lr}
		\addlinespace
		\toprule
		\multicolumn{2}{c}{\textbf{Path Parameters}}\\ \midrule
		AoA & $30^{\circ}$, $150^{\circ}$, $-45^{\circ}$\\
		EoA & $35^{\circ}$, $50^{\circ}$, $75^{\circ}$\\
		normalised delay & $\frac{1}{9}$, $\frac{2}{9}$, $\frac{4}{9}$\\
		path polarisation & horizontal, horizontal, vertical\\
		normalised path power & 0 dB, -2 dB, -3 dB \\
		\# of samples & 128\\
		\midrule
	\end{tabular}
	\caption{Path Parameters}
	\label{tab:PathParameters}
\end{table}

We can see in every plot, that the RMSE shrinks with increasing SNR. Hence we assume the estimators as consistent.
Furthermore, the RMSE of each parameter is in an acceptable range. This points out, that the paths could be resolved well.
It is obvious for low SNR values, that the RMSE of elevation, delay and normalised power is smaller for the CML, compared to the CCR. For higher SNR values, no difference between the RMSEs of the cost functions can be determined. Therefore we assume, that both algorithms behave asymptotical equal for high SNR values. Based on this fact, we cannot select the appropriate cost function from the carried out simulations. Other criteria like the number of optimisation iterations or stability of the path number estimation have to be selected. Such criteria are behind the scope of this paper.
\section{Conclusion}
\label{sec:Conclusion}

In this paper, the problem of radio channel parameter estimation for an unknown transmitter was investigated. Based on a ray-optical model, sufficient parameters were introduced and the accessibility of these parameters was clarified. We proposed two cost functions, which overcome the issue of the unknown transmitter signal in different ways. A Monte-Carlo simulation showed, that these cost functions are applicable for parameter estimation. A preference for one of the presented cost functions based on the carried out simulations could not be given.
\section*{Acknowledgment}
This work is part of the EiLT-project \cite{EiLT2013}, funded by the German Federal Ministry of Education and Research (BMBF).

\bibliographystyle{IEEEtran}
\bibliography{./bibliography}

\end{document}